\documentclass[journal,final, twocolumn]{IEEEtran}
\usepackage{cite}
\usepackage{amsmath,lipsum}
\usepackage{graphicx}
\usepackage{color}
\begin{document}
\title{A Five-Freedom Active Damping and Alignment Device Used in the Joule Balance}
%
%
%

\author{Jinxin Xu, Qiang You, Zhonghua Zhang, Zhengkun Li and Shisong Li,~\IEEEmembership{Member,~IEEE}
\thanks{This work is supported by the China National Natural Science Foundation (Grant Nos.91536224, 51507088) and the China National Key Research and Development Plan (Grant No. 2016YFF0200102).
}
\thanks{J. Xu, Q. You are with Tsinghua University, Beijing 100084, China.

Z. Zhang and Z. Li are with the National Institute of Metrology (NIM), Beijing 100029, China and the Key Laboratory for the Electrical Quantum Standard of AQSIQ, Beijing 100029, China.

S. Li is currently with the International Bureau of Weights and Measures (BIPM), Pavillon de Breteuil, F-92312 S\`{e}vres Cedex, France. E-mail:leeshisong@sina.com.}}

%
%

\markboth{IEEE Trans. Instrum. Meas., CPEM2016, \today}%
{}
%



\maketitle

\begin{abstract}
Damping devices are necessary for suppressing the undesired coil motions in the watt/joule balance. In this paper, an active electromagnetic damping device, located outside the main magnet, is introduced in the joule balance project. The presented damping device can be used in both dynamic and static measurement modes. With the feedback from a detection system, five degrees of freedom of the coil, i.e. the horizontal displacement $x$, $y$ and the rotation angles $\theta_x$, $\theta_y$, $\theta_z$, can be controlled by the active damping device. Hence, two functions, i.e. suppressing the undesired coil motions and reducing the misalignment error, can be realized with this active damping device. The principle, construction and performance of the proposed active damping device are presented.
\end{abstract}


%
\IEEEpeerreviewmaketitle

\section{Introduction}
%
%
%
%
Several national metrology institutes (NMIs) across the world are working hard on redefining one of the seven SI base units, the kilogram (kg), in terms of the Planck constant $h$, by means of either the watt/joule balance [1-9] or the X-ray crystal density (XRCD) method [10]. The watt balance, which was proposed by Kibble in 1975 [11], has been adopted by the majority of NMIs. The National Institute of Metrology (NIM, China) is focusing on the joule balance, which can be seen as an alternative realization of the watt balance [12].

The unwanted coil motions in the watt/joule balance, e.g., horizontal and rotational movements, will lower the signal to noise ratio in the measurement. With these motions, the measurement would take a much longer time in order to achieve the type A relative uncertainty at an order of $10^{-8}$. Moreover, the noticeable undesired coil motions will introduce a systematic bias [13]. Therefore, damping devices are necessary to be employed to suppress these undesired coil motions in the watt/joule balance. In watt balances, e.g., NIST-4 at the National Institute of Standards and Technology (NIST, USA) [14], an active damping system has been used during the measurement. The damping coils in such systems are conventionally located on the framework of the suspended coil. To avoid additional force in the weighing mode and unwanted induced voltage in the velocity mode, the damping device is turned off during both measurement modes. During the dynamic measurement mode, the undesired coil motions, however, is not taken care of. In the joule balance at NIM to suppress unwanted motions during the measurement without introducing any flux into the main magnet, a novel active electromagnetic damping device, located outside the main magnet, is designed and used. The merit of the system is that the field of the damping device has no effect on the main field, and hence can be used in both the dynamic and static measurement modes. With the feedback from a detection system, five degrees of freedom of the coil, i.e. the horizontal displacement $x$, $y$ and the rotation angles $\theta_x$, $\theta_y$, $\theta_z$, can be controlled by this device. Two functions, i.e. suppressing the undesired coil motions and reducing the misalignment error, then can be realized in the measurements. The original idea was presented in [15]. In this paper, the principle, construction and performance of the active electromagnetic damping device are presented.

The rest of this article is organized as follows: the principle of the joule balance is introduced in section 2; the principle and construction of the active damping device are presented in section 3; the performance of the active damping device in the joule balance is presented in section 4; some potential systematic effects using the active damping device, the vertical force and the flux leakage at the mass weighing position, are discussed in section 5.

\section{Principles of Joule Balance}
The joule balance is the integral of the watt balance. The mathematical details are presented in [12]. The equation is expressed as
\begin{equation}
\int_LF\cdot {\rm d}l+\int_L\tau\cdot {\rm d}\theta=I[\psi({\rm B})-\psi({\rm A})],
\end{equation}
where $F$ denotes the magnetic force, $\tau$ the torque relative to the mass center of the coil, $L$ the vector trajectory when the coil is moved from position A to B, $I$ the current through the coil, $\psi({\rm B})-\psi({\rm A})$ the flux linkage difference of the position B and A.

On the right side of equation (1), the product of the flux linkage difference $\psi({\rm B})-\psi({\rm A})$ and the current $I$ is the magnetic energy change. On the left side of equation (1), the two integral terms are the work done by the magnetic force and torque. They can be expressed as
\begin{equation}
\begin{aligned}
w_{\rm AB}=\int_LF_z {\rm d}z+\int_LF_x {\rm d}x+\int_LF_y {\rm d}y\\+\int_L\tau_z {\rm d}\theta_z+\int_L\tau_x {\rm d}\theta_x+\int_L\tau_y {\rm d}\theta_y.
\end{aligned}
\end{equation}

\begin{figure}[!t]
\centering
\includegraphics[width=0.48\textwidth]{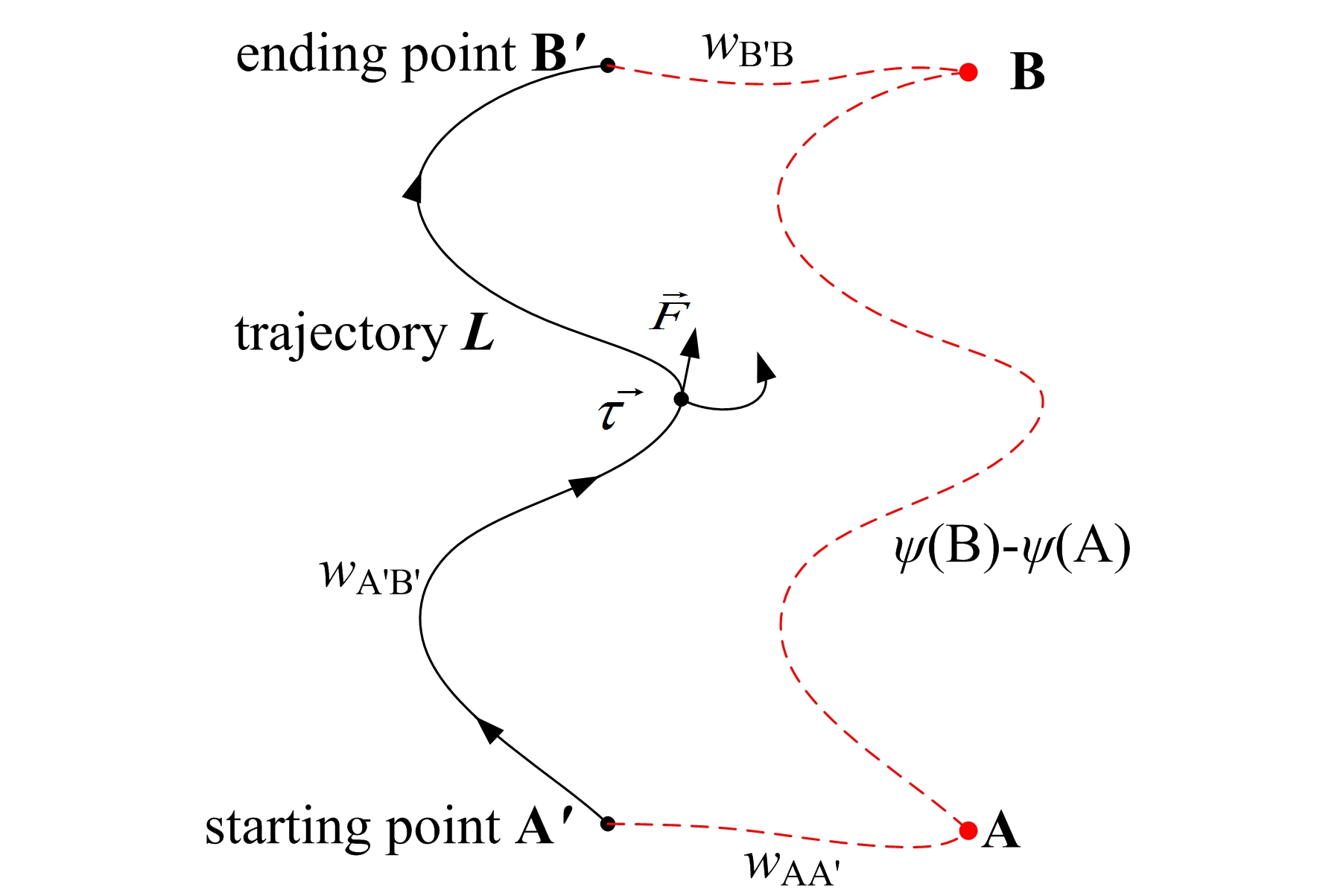}
\caption{The schematic diagram of the misalignment errors.}
\label{fig1}
\end{figure}

Note that in equation (2) only the vertical force, i.e. $F_z$, can be measured precisely. The integral terms of the horizontal forces $F_x$, $F_y$ and torques $\tau_x$, $\tau_y$, $\tau_z$ are parts of the misalignment errors in the joule balance. Other parts of the misalignment errors come from the position change of the coil between the two measurement modes in the joule balance. In equation (1), it is assumed that the positions A and B in the measurement of the flux linkage difference are the same as the starting point and ending point of the integral trajectory $L$. However, it is not the case in the actual measurement. As shown in Fig. 1, when current pass through the coil, the positions A and B will actually shift to A' and B'. Then, equation (1) can be rewritten as
\begin{equation}
w_{\rm AA'}+w_{\rm A'B'}+w_{\rm B'B}=I[\psi({\rm B})-\psi({\rm A})],
\end{equation}
where $w_{\rm AA'}$, $w_{\rm A'B'}$, $w_{\rm B'B}$ are the work done by the magnetic force and torque when the coil is respectively moved from position A to A', A' to B' and B' to B. It can be seen from equation (3) that the misalignment error includes three parts: $w_{\rm AA'}$, $w_{\rm B'B}$ and the integral terms of the horizontal forces and torques in $w_{\rm A'B'}$.

The active damping device used in the joule balance is able to produce auxiliary horizontal forces and torques on the coil to change the position of the coil. Therefore, with the feedback from the detection system of the five directions of displacement, the position of the coil can be easily controlled. As a result, the positions A and B can be adjusted to be the same as A' and B' by the feedback of the damping device. In the meanwhile, the horizontal displacement $x$, $y$ and the rotation angles $\theta_x$, $\theta_y$, $\theta_z$, can be kept unvaried when the coil is moved along the vector trajectory $L$ in the weighting mode of the joule balance. These two improvements will significantly reduce the misalignment error. For an ideal case, i.e. the alignment is adjusted to be good enough, only the work done by the vertical force in $w_{\rm A'B'}$ is left on the left side of equation (3) and other terms can be ignored. Then equation (3) can be rewritten as
\begin{equation}
\int_{\rm A'}^{\rm B'}F_z{\rm d}z=I[\psi({\rm B})-\psi({\rm A})].
\end{equation}

\section{Principle and Construction of the Damping Device}

\subsection{The practical structure}
Fig. 2 shows the practical structure of the active damping device used in the joule balance.The outer yokes 1, 2, 3 and the inner yokes 8, 9,10 are made of soft iron with high permeability. The component 4 shown in Fig. 2(a) is made of aluminum, for mechanically supporting three inner yokes. The component 17 in Fig. 2(c) and Fig. 2(d) is the framework of the auxiliary coils which is made up of polysulfone.

Components 5, 6, 7, 11, 12 and 13 are permanent magnets magnetized along the $z$ axis. It should be noted that the magnetization direction of the three permanent magnets 5, 12, 13 is opposite to the magnetization direction of the other three permanent magnets 6, 7, 11. As a result, in this magnetic circuit, the magnetic flux in the air gap originates from the yoke 8 and enters the yokes 9, 10.

\begin{figure}[!t]
\centering
\includegraphics[width=0.48\textwidth]{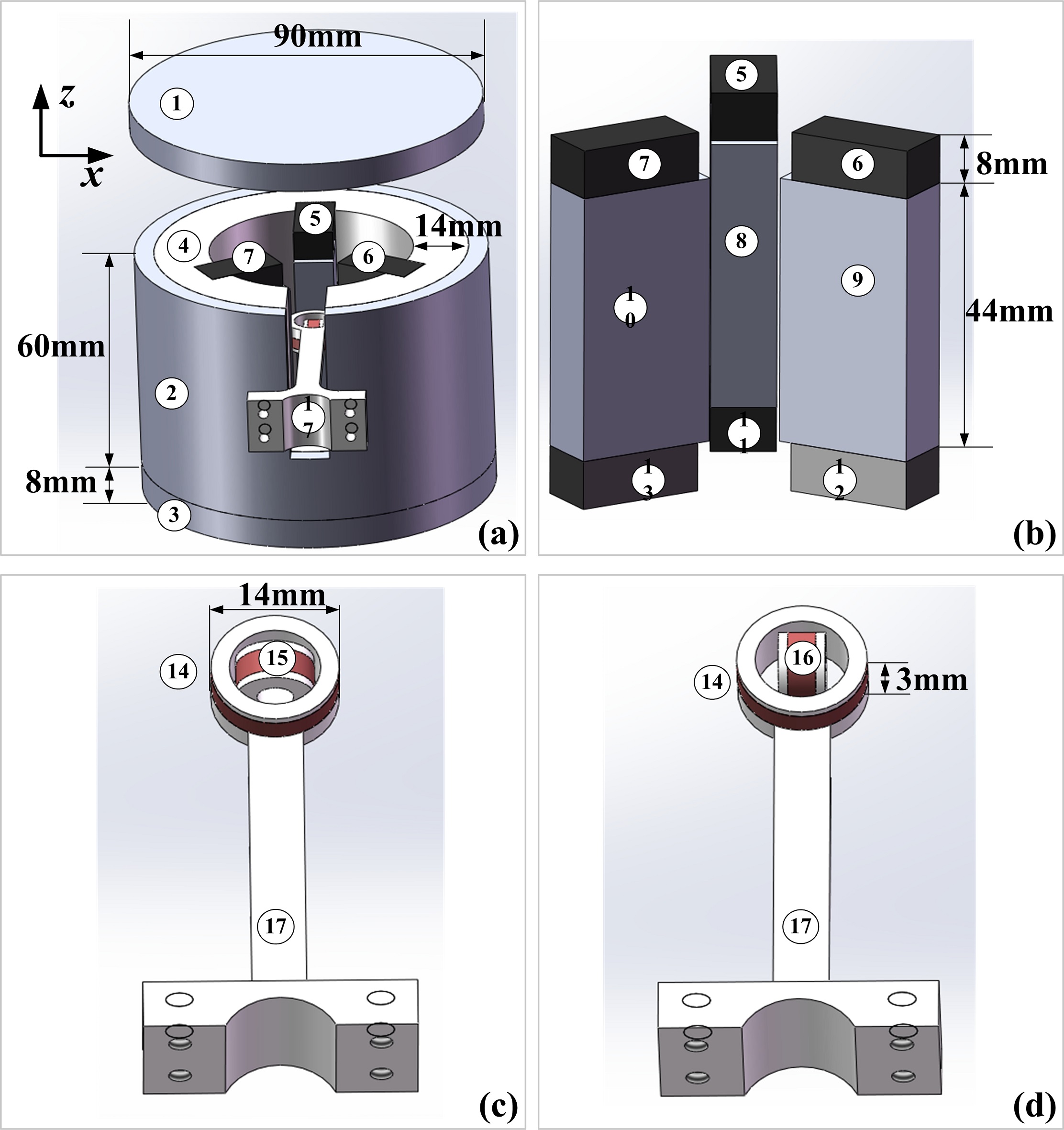}
\caption{Construction of the damping device. (a) The overall structure of the damping device with the top cover 1 open. (b) The permanent magnets (5, 6, 7, 11,12, 13) and the inner yokes (8, 9, 10). (c) Combination of auxiliary coils 14 and 15. (d) Combination of auxiliary coils 14 and 16.}
\label{fig2}
\end{figure}

Components 14, 15 and 16 are three orthogonal auxiliary coils. Fig. 2(c) and Fig. 2(d) show two different combinations of the auxiliary coils. Coils 14 and 15 are glued together in Fig. 2(c) and coils 14 and 16 are glued together in Fig. 2(d). Since there are considerable magnetic gradients in $x$, $y$ and the azimuth $\theta$, torques and forces will be produced on the auxiliary coils when current passes through the auxiliary coils.

The cross sectional view of the inner yokes and the auxiliary coils in the active damping device is shown in Fig. 3. The dotted lines denote the magnetic flux in the air gap. The coil 14 in the magnetic field shown in Fig. 3(a) will produce a torque $\tau_x$ around $x$ axis, relative to the mass center. Coil 15 in Fig. 3(b) will produce a force $F_y$ along the $y$ axis and coil 16 in Fig. 3(c) will produce a force $F_x$ along the $x$ axis. The forces and torques of the three coils along other directions are theoretically small.

The schematic diagram of the damping device is shown in Fig. 3(d), containing 3 damping segments. The coil is suspended from the spider by three rods. Each damping segment is fixed on a support with the auxiliary coils mechanically connected to the coil spider rod. In two of them, the combination of the auxiliary coils 14 and 15 shown in Fig. 2(c) is used. In the other one, the combination of the auxiliary coils 14 and 16 shown in Fig. 2(d) is used. Hence, the torque produced by coil 14 in all the three damping segments is used to adjust the rotation angle $\theta_x$, $\theta_y$ of the suspended coil. The force of coil 15 in two segments points to the center of the suspended coil and can be applied to adjust the horizontal movement $x$, $y$. The force of coil 16 in the third segment is along the tangential of the suspended coil and can be employed to adjust the rotation along $z$ axis, i.e. $\theta_z$.

\begin{figure}[!t]
\centering
\includegraphics[width=0.48\textwidth]{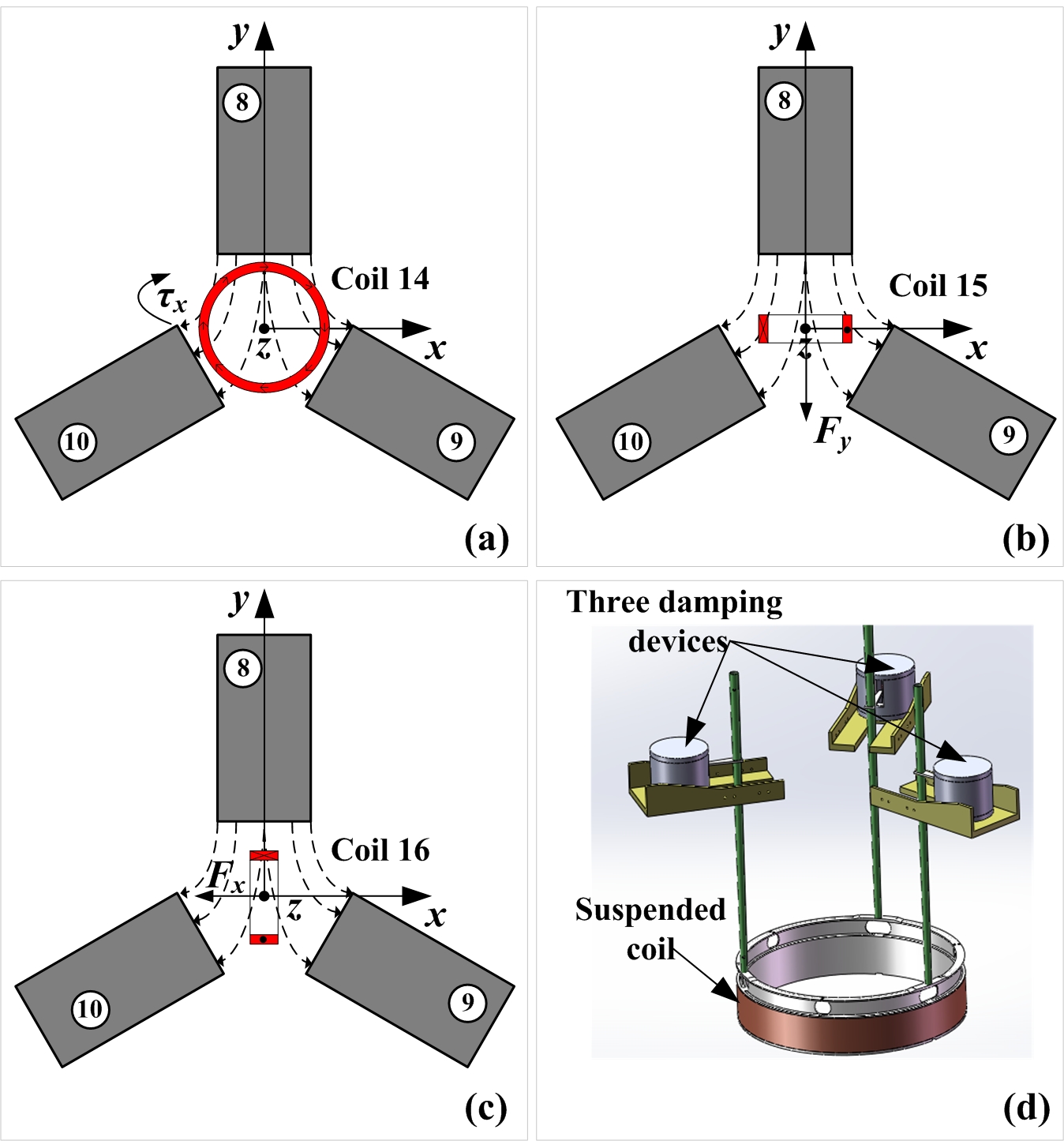}
\caption{The cross sectional view of the inner yokes and the auxiliary coils. (a) The torque of coil 14. (b) The force of coil 15. (c) The force of coil 16. (d) The schematic diagram of three damping devices.}
\label{fig3}
\end{figure}

\begin{figure}[!t]
\centering
\includegraphics[width=0.5\textwidth]{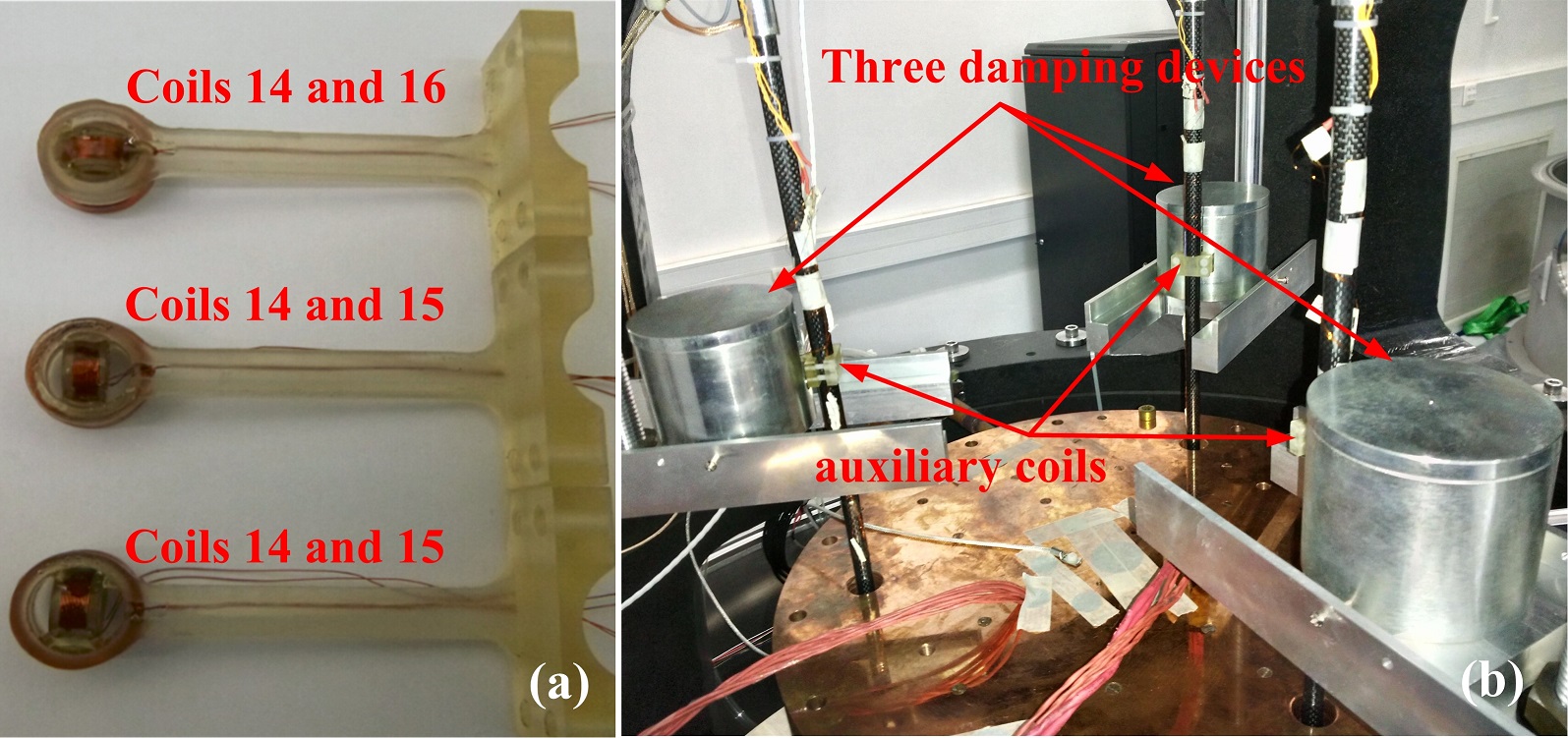}
\caption{(a) The auxiliary coils installed on the suspended system. (b) Mechanical assembling of the damping device.}
\label{fig4}
\end{figure} 	

The auxiliary coils are connected to the current sources through the suspension system by fine wires called hairsprings. To reduce the number of the hairsprings and keep the symmetry of the suspended coil, six auxiliary coils are installed in the practical structure instead of nine in the initial design [15]. Fig. 4(a) shows the six auxiliary coils installed on the suspended system. In theory, to suppress the motions of the five degrees of freedom,  the minimum number of auxiliary coils is five. Hence, only two of the three coils 14 are connected to the current sources in practice. The three damping devices are placed at angles of 120 degrees around the magnet in the joule balance as shown in Fig. 4(b). The pallets of the damping devices are fixed with respect to the marble shelf in the joule balance which will be kept static.

In the joule balance, the suspended coil is kept static while the magnet is moved by a linear translation stage. For the active damping device used in the joule balance, the auxiliary coils are always kept static and do not require too much moving space. If such a damping device is used in the velocity mode of watt balances, the length of the inner yokes must be long enough for the movement of the auxiliary coils.

\subsection{The feedback control}
The entire control circuit is shown in Fig. 5. The PID controller is realized with a LabVIEW program. Five channels of a high-speed analog output (NI 6733) are used as the voltage control of the current sources. With an external voltage reference, the range of the analog output voltage is $\pm$1\,V. The circuit of a current source is shown in Fig. 6, performing the $U/I$ converter. A 10\,$\Omega$ four-terminal resistor with a low temperature coefficient ($<$1 ppm/$^{\circ}$C) is used as the sense resistor in the current source. Hence, the range of the output current is $\pm$100\,mA.

\begin{figure}[!t]
\centering
\includegraphics[width=0.47\textwidth]{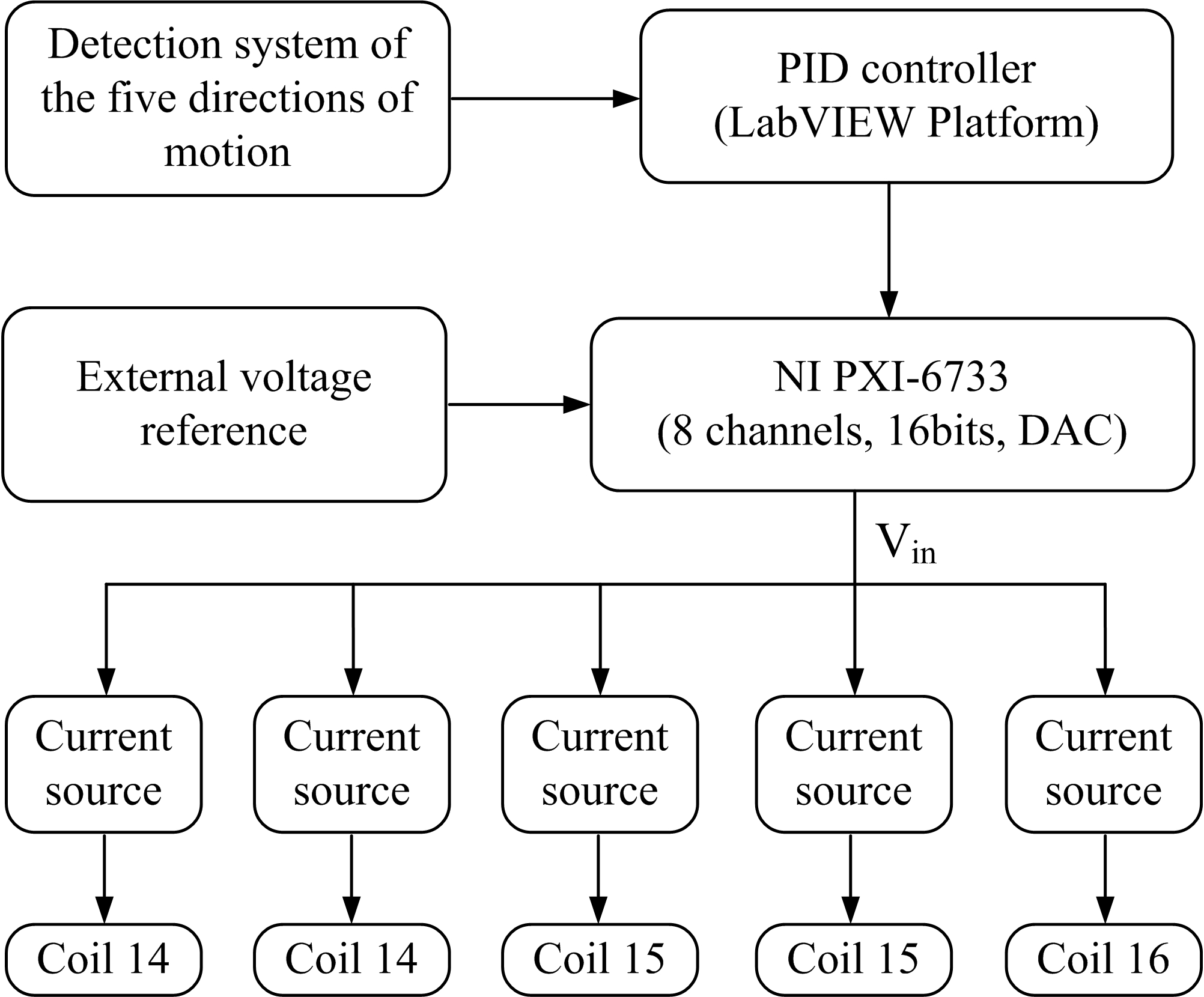}
\caption{The block diagram of the active control circuit.}
\label{fig5}
\end{figure}

\begin{figure}[!t]
\centering
\includegraphics[width=0.48\textwidth]{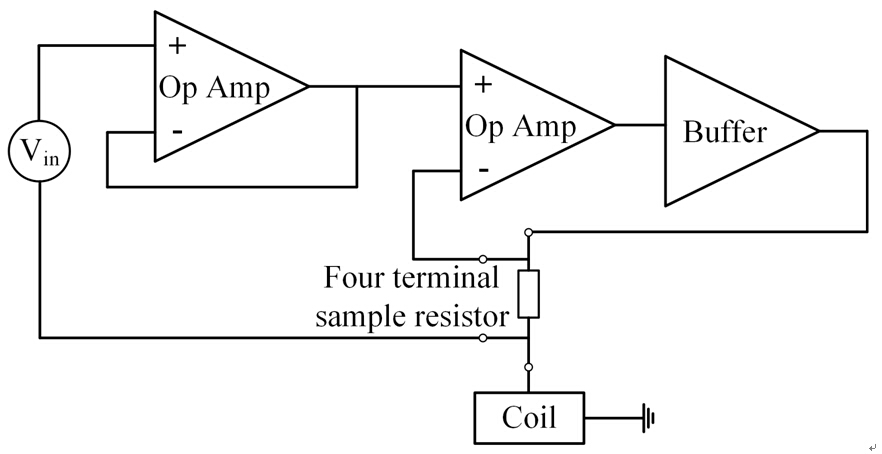}
\caption{The circuit of the current source.}
\label{fig6}
\end{figure}

The detection system of the relative position between the coil and the main magnet is composed of a laser interferometer and position sensitive devices (PSD). The rotation angles $\theta_x$, $\theta_y$ are obtained with the laser interferometer. The resolution of $\theta_x$ and $\theta_y$ is less than 1\,$\mu$rad. The horizontal displacements $x$, $y$ and rotation angle $\theta_z$ are obtained from four PSDs. The resolution in $x$ and $y$ is about 2\,$\mu$m and the resolution of $\theta_z$ is about 5\,$\mu$rad.

\begin{figure}[!t]
\centering
\includegraphics[width=0.455\textwidth]{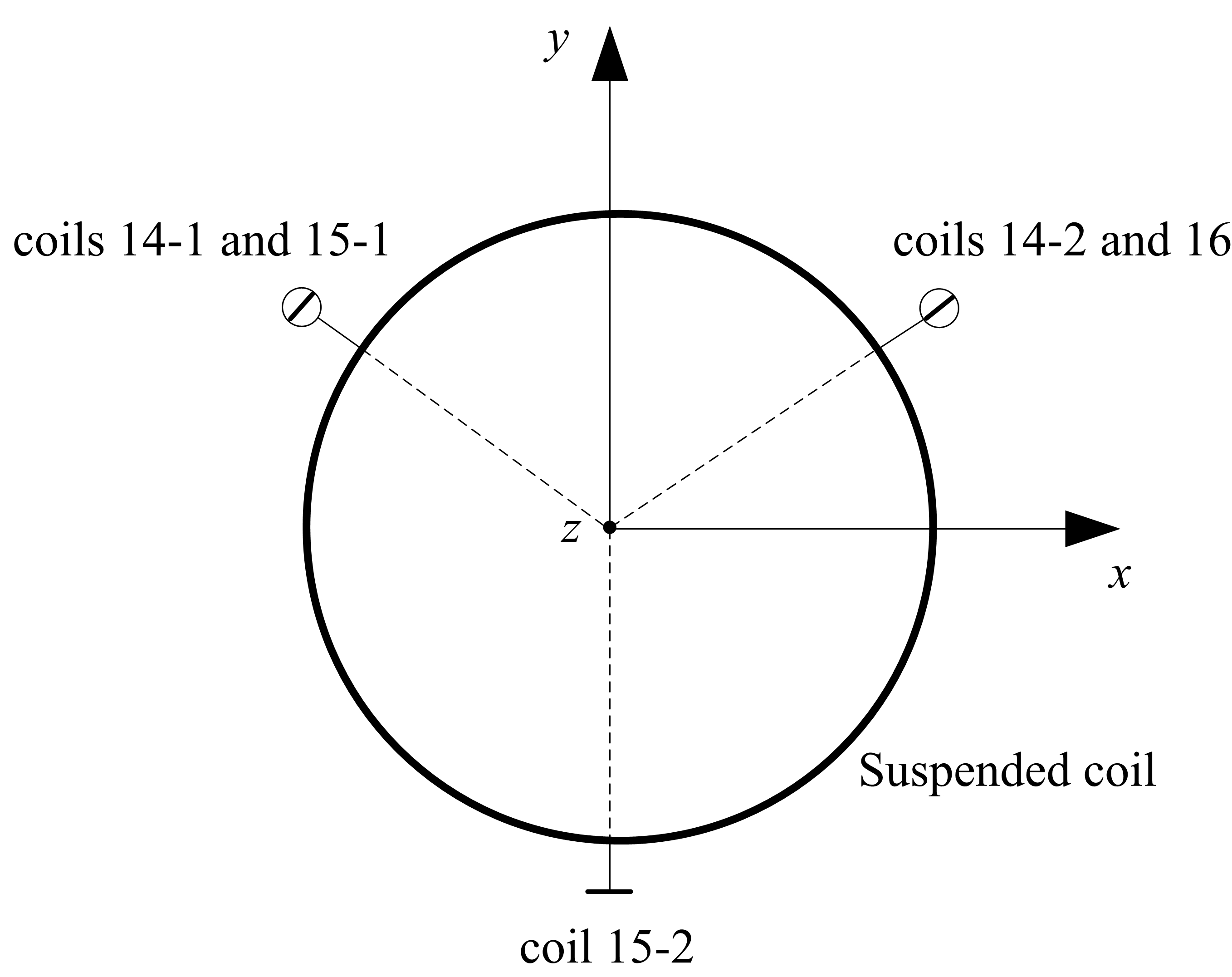}
\caption{The setup of auxiliary coils and the coordinate axis of the detection system.}
\label{fig7}
\end{figure}

The five auxiliary coils and the coordinate axis of the detection system are shown in Fig. 7. The large circle denotes the suspended coil in joule balance. Five auxiliary coils are set concentric to the suspended coil with every 120 degrees. Since the horizontal displacement and rotation angle of the suspended coil are in a small range, the relationship between the detection and the auxiliary coil current is close to be linear. In order to check the dependence between the coil motion and the current through the auxiliary coils, the transfer matrix $M$ is measured by applying a 50\,mA current individually to each of the five auxiliary coils. The measurement is written as
\begin{equation}
\begin{aligned}
&~~~\left[
\begin{array}{ccc}
{\rm d}x\\
{\rm d}y\\
{\rm d}\theta_x\\
{\rm d}\theta_y\\
{\rm d}\theta_z
\end{array}\right]
=M\left[
\begin{array}{ccc}
I_{16}\\
I_{15-2}\\
I_{15-1}\\
I_{14-2}\\
I_{14-1}
\end{array}\right]\\
&=\left[\begin{array}{rrrrr}
0.12    &{\bf0.32}  & {\bf-1.04}  & -0.20   & 0.16\\
    0.16   &{\bf-1.44}   & {\bf0.88}   &-0.24  & -0.24\\
    0.10    &0.12    &0.16    &{\bf1.78}    &{\bf1.30}\\
    0.04    &0.10   &-0.08   &{\bf-1.16}    &{\bf1.78}\\
    {\bf6.30}    &0.20   &-0.30 &        0.00      &   0.00
\end{array}\right]
\left[
\begin{array}{ccc}
I_{16}\\
I_{15-2}\\
I_{15-1}\\
I_{14-2}\\
I_{14-1}
\end{array}\right],
\end{aligned}
\end{equation}
where ${\rm d}x$, ${\rm d}y$, ${\rm d}\theta_x$, ${\rm d}\theta_y$ and ${\rm d}\theta_z$ are the variations of the detection system with units of $\mu$m, $\mu$m, $\mu$rad, $\mu$rad, $\mu$rad respectively, $I_{16}$, $I_{15-2}$, $I_{15-1}$, $I_{14-2}$ and $I_{14-1}$ the currents passing through the five auxiliary coils with unit of mA.

The determined matrix $M$ in equation (5) matches the analysis of the above section: The horizontal displacement $x$, $y$ is mainly controlled by coils 15-1 and 15-2. The rotation angle $\theta_x$, $\theta_y$ is mainly adjusted by coils 14-1 and 14-2. The rotation angles $\theta_z$ is mainly controlled by coil 16. It can be seen that due to some imperfection of the magnetic field and the misalignment of five auxiliary coils, there are some weak couplings between different channels. In the control, we should use the inverse of the transfer matrix to decouple the correlation of different loops, i.e.
\begin{equation}
\begin{aligned}
&~~~\left[
\begin{array}{ccc}
I_{16}\\
I_{15-2}\\
I_{15-1}\\
I_{14-2}\\
I_{14-1}
\end{array}\right]
=M^{-1}\left[
\begin{array}{ccc}
{\rm d}x\\
{\rm d}y\\
{\rm d}\theta_x\\
{\rm d}\theta_y\\
{\rm d}\theta_z
\end{array}\right]\\
&=\left[\begin{array}{rrrrr}
-0.03	&	0.02	&	0.00	&	0.00	&	0.16	\\
-0.78	&	-0.87	&	-0.16	&	0.07	&	0.04	\\
-1.22	&	-0.26	&	-0.08	&	0.14	&	0.03	\\
0.12	&	0.04	&	0.39	&	-0.29	&	-0.01	\\
0.07	&	0.06	&	0.26	&	0.37	&	-0.01	
\end{array}\right]
\left[
\begin{array}{ccc}
{\rm d}x\\
{\rm d}y\\
{\rm d}\theta_x\\
{\rm d}\theta_y\\
{\rm d}\theta_z
\end{array}\right].
\end{aligned}
\end{equation}

With a known difference of the detection and the target, the output voltage of auxiliary coils will be determined by both the PID controller and the inverse matrix $M^{-1}$. Note that since the response of auxiliary coils can easily excite the suspension mechanism, fast and large feedback currents should be avoided.

\section{Experimental Results}
\subsection{Damping performance}
When the suspended coil receives external shocks such as loading of the test mass, ground vibration, etc., the energy blue transfer out through the suspension system is very slow due to a high $Q$ factor. The suspended coil will swing for a long time without damping. Fig. 8 shows the motions of the suspended coil when the mass is put on the mass pan with and without the proposed damper. Without any damping device, the maximal motion amplitudes of the five degrees of freedom $x$, $y$, $\theta_x$, $\theta_y$, $\theta_z$ are 250\,$\mu$m, 150\,$\mu$m, 500\,$\mu$rad, 550\,$\mu$rad, 300\,$\mu$rad respectively. After about 5 minutes, the motion amplitudes will decrease to 10\,$\mu$m, 10\,$\mu$m, 15\,$\mu$rad, 15\,$\mu$rad, 50\,$\mu$rad.The damping factor is about 0.11\,kg$\cdot$s$^{-1}$. The steady state deviations between the mass on state and the mass off state of the five degrees of freedom are 10\,$\mu$m, 35\,$\mu$m, 50\,$\mu$rad, 20\,$\mu$rad, 30\,$\mu$rad respectively.

When the active damping device is used, the motions of the suspended coil will decay rapidly. As shown in Fig. 8, $x$, $y$, $\theta_x$, $\theta_y$, $\theta_z$ will reduce to 8\,$\mu$m, 8\,$\mu$m, 10\,$\mu$rad, 10\,$\mu$rad, 20\,$\mu$rad in 30 seconds. The damping factor is about 1.1\,kg$\cdot$s$^{-1}$, which is about 10 times larger than that without the damper. The steady state deviations between the mass on state and the mass off state of the five degrees of freedom are less than 2\,$\mu$m, 2\,$\mu$m, 1\,$\mu$rad, 1\,$\mu$rad, 5\,$\mu$rad.

\begin{figure}[!t]
\centering
\includegraphics[width=0.46\textwidth]{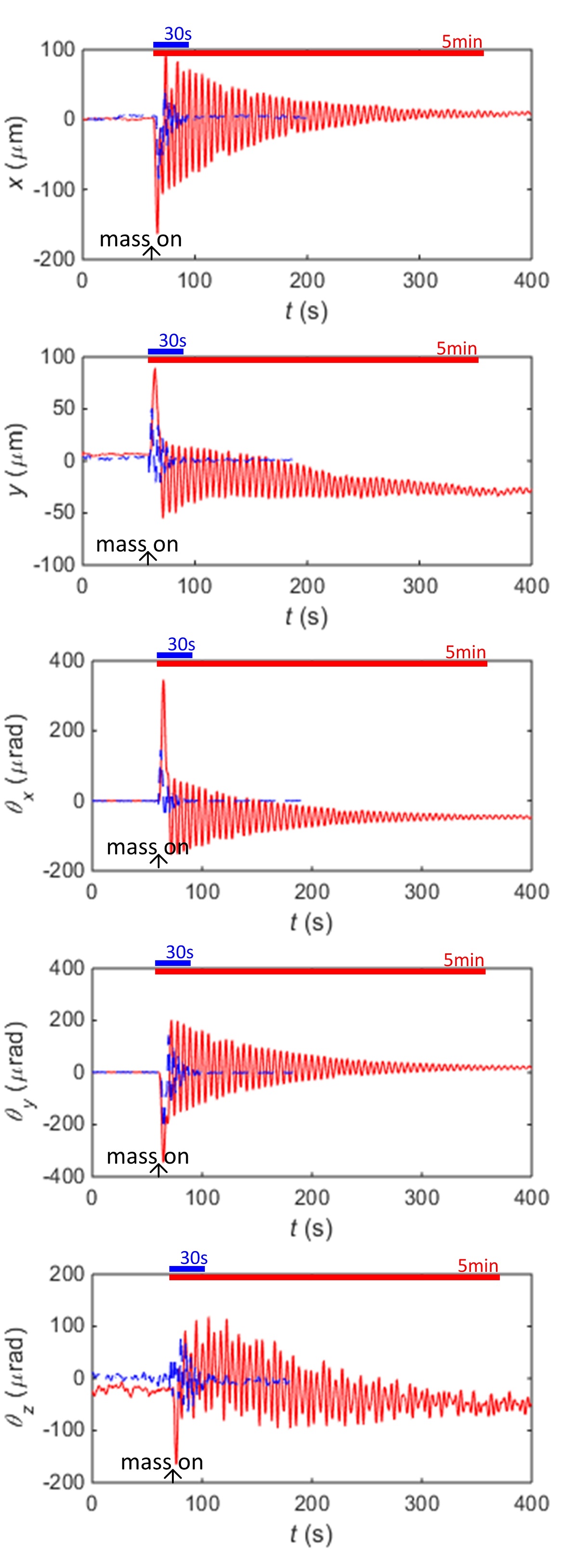}
\caption{Coil motions when test mass is put on the mass pan with and without the proposed damper. The red lines (--) are without the damper while the blue dash lines (- -) are with the damper.}
\label{fig7}
\end{figure}

\subsection{Reduction of the current dependence of the coil position}
When current passes through the suspended coil, the positions A and B in the flux linkage difference measurement will change to A' and B', yielding the errors $w_{\rm AA'}$ and $w_{\rm B'B}$. To generate a 5\,N magnetic force, the suspended coil current is about 7.5\,mA. Fig. 9 shows the changes of the five degrees of freedom $x$, $y$, $\theta_x$, $\theta_y$, $\theta_z$ when the suspended coil current is reduced from 7.5\,mA to 0\,mA slowly. Without the active damping device, the magnitude changes are 25\,$\mu$m, 5\,$\mu$m, 15\,$\mu$rad, 15\,$\mu$rad, 70\,$\mu$rad respectively. Due to the restriction of the narrow air gap and the laser alignment in the present construction of the joule balance, it is very difficult to further adjust the concentricity and horizontality of the suspended coil to reduce the magnitudes of the coil position change.

To reduce the errors $w_{\rm AA'}$ and $w_{\rm B'B}$, the positions of A and B can be adjusted to be the same as A' and B' by the active damping device. As shown in Fig. 9, when the active damping device is added, the magnitude changes of $x$, $y$, $\theta_x$, $\theta_y$, $\theta_z$ are less than 2\,$\mu$m, 2\,$\mu$m, 1\,$\mu$rad, 1\,$\mu$rad, 5\,$\mu$rad respectively, which on average is one magnitude improved than that without the electromagnetic damper.

\begin{figure}[!t]
\centering
\includegraphics[width=0.465\textwidth]{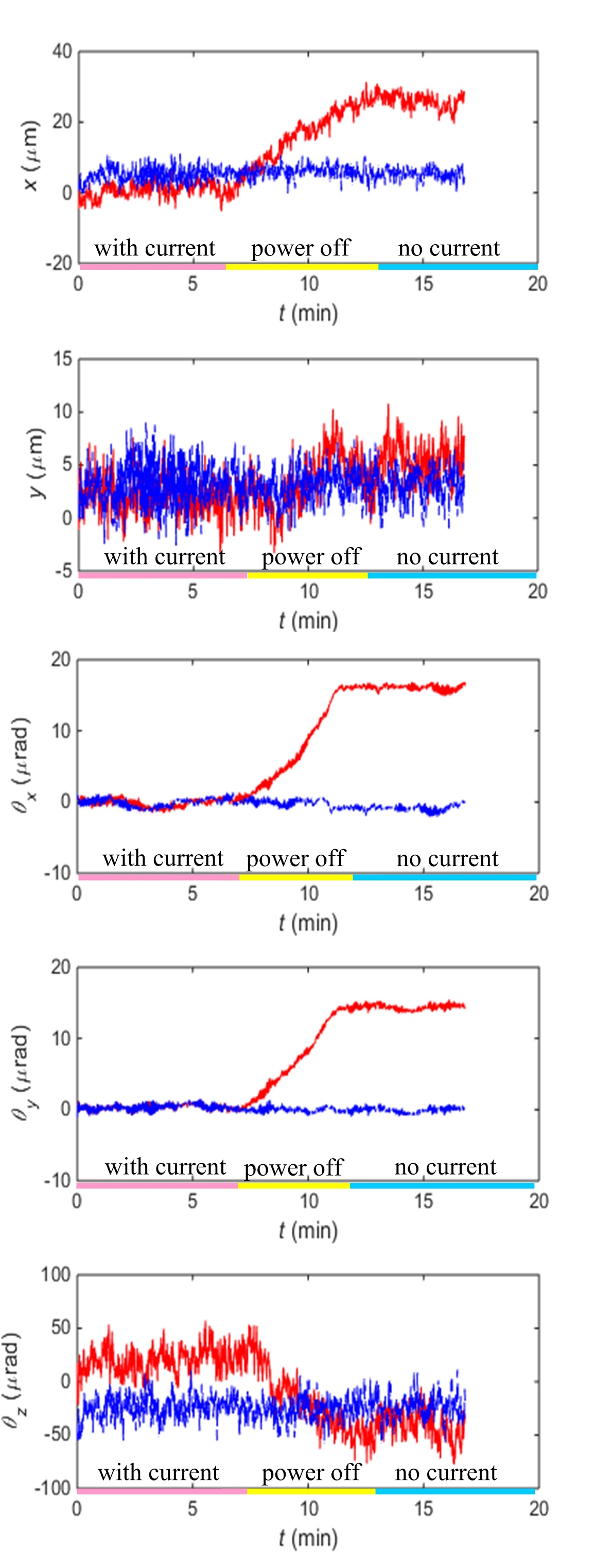}
\caption{Coil motions when the coil current is reduced from 7.5\,mA to 0\,mA slowly. The red lines (--) are without the damper while the blue dash lines (- -) are with the damper.}
\label{fig8}
\end{figure}

\begin{figure}[!t]
\centering
\includegraphics[width=0.468\textwidth]{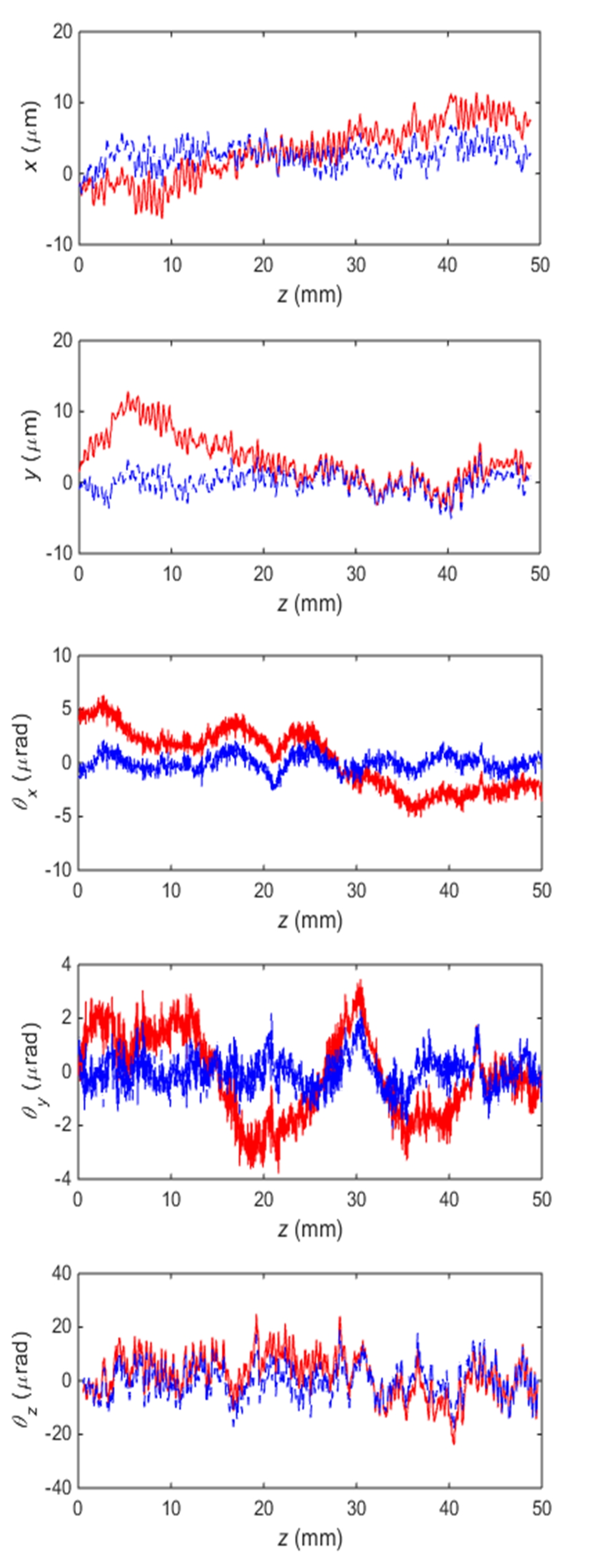}
\caption{Coil motions with different $z$ during the movement of the magnet. The red lines (--) are without the damper while the blue dash lines (- -) are with the damper.}
\label{fig9}
\end{figure}

\subsection{Improvement of the vertical movement}
In both the weighing and dynamic modes of the joule balance, the magnet is moved in the vertical direction $z$ by a translation stage [17]. However, due to the distortion of the guide rail in the translation stage and other mechanical excitations, the other five degrees of freedom $x$, $y$, $\theta_x$, $\theta_y$, $\theta_z$ will also change with the movement of the magnet. Therefore, the horizontal forces $F_x$, $F_y$ and torques $\tau_x$, $\tau_y$, $\tau_z$ will contribute to in the misalignment errors.

Fig. 10 shows the value of the five degrees of freedom $x$, $y$, $\theta_x$, $\theta_y$, $\theta_z$ with different $z$ during the movement of the magnet. Without the active damping device, the peak-to-peak values of the five degrees of freedom are 15\,$\mu$m, 15\,$\mu$m, 10\,$\mu$rad, 8\,$\mu$rad, 40\,$\mu$rad respectively. When the active damping device is added during the movement, the unwanted motions of the magnet can be compensated by the motions of the suspended coil. Hence, the changes of the relative value of $x$, $y$, $\theta_x$, $\theta_y$, $\theta_z$ can be reduced. As shown in Fig. 10, the peak-to-peak values of the five degrees of freedom are reduced to 6\,$\mu$m, 6\,$\mu$m, 4\,$\mu$rad, 4\,$\mu$rad, 25\,$\mu$rad with the active damping device. The measurement shows that all curves of the five degrees of freedom are nearly flat. Therefore, the misalignment errors, resulting from the work done by the horizontal forces and torques in $w_{\rm A'B'}$, are greatly reduced.

\section{Consideration of Potential Systematic Effects}

\subsection{The vertical force}
When the active electromagnetic damping device is used in the weighing mode of the joule balance, any additional vertical force caused by the five auxiliary coils should be considered. Here the mass comparator has been used to measure the vertical force. With 100\,mA current in the auxiliary coils, the measurement results are shown in Table I. As discussed in [15], the vertical forces of coil 16 and 15 are very small when compared with that of coil 14. Since the working current in the auxiliary coils is much less than 100\,mA in the weighing mode, the vertical forces will be much smaller than the results in Table I.

\begin{table}[!t]
\centering
\caption{The unwanted vertical force of the five auxiliary coils with 100\,mA excitation.}
\begin{tabular}{ccc}
\hline
Coil No. & Vertical force /mg & uncertainty /mg\\
\hline
 16 & 0.08 &0.01\\
 15-1 & 0.06&0.01\\
 15-2& 0.06&0.01\\
 14-1& 1.04&0.01\\
 14-2 & 0.85&0.01\\
\hline
\end{tabular}
\end{table}

When the force of the suspended coil is measured in the weighing mode, the suspended coil is static and the current in the auxiliary coils of the active damping device is nearly constant.  As the magnetic force is proportional to the current, the additional vertical force of the auxiliary coil can be determined by measuring the current in the auxiliary coil. Note that in this measurement, the uncertainty of the mass comparator is about 0.01\,mg. A correction may then be made and the effect of the unwanted vertical forces can be reduced to several parts in $10^{8}$.

\subsection{Magnetic flux leakage}
The distance between the active damping devices and the test mass is about 160\,mm. Hence, the leakage magnetic field of the active damping devices should be considered. The interaction between the magnetic field and the mass is discussed in [16]. The equation of the vertical force on the mass with a volume susceptibility $\chi$ and permanent magnetization $M$ is given by [18]
\begin{equation}
F_z=-\frac{\mu_0}{2}\frac{\partial}{\partial z}\int\chi H\cdot H{\rm d}V-\mu_0\frac{\partial}{\partial z}\int M\cdot H{\rm d}V.
\end{equation}

The magnetic field at the location of the mass is measured by a gauss meter. Fig. 11 shows the absolute magnitude of the magnetic field as a function of the distance from the top surface of the mass pan. Note that although the measured field is comparable to the earth's magnetic field, it produces a much larger field gradient, about 1.3\,$\mu$T/mm. The employed 500\,g test mass is about 50\,mm in height and 40\,mm in diameter. The magnetic susceptibility $\chi$ of the mass material is less than $6\times10^{-4}$ and the permanent magnetization $M$ is about 0.1\,A/m. With the method used in [16], the calculated magnetic force of the mass is less than 10\,$\mu$g. Hence, the relative systematic error caused by the leakage magnetic field is less than $2\times10^{-8}$.

\begin{figure}[!t]
\centering
\includegraphics[width=0.46\textwidth]{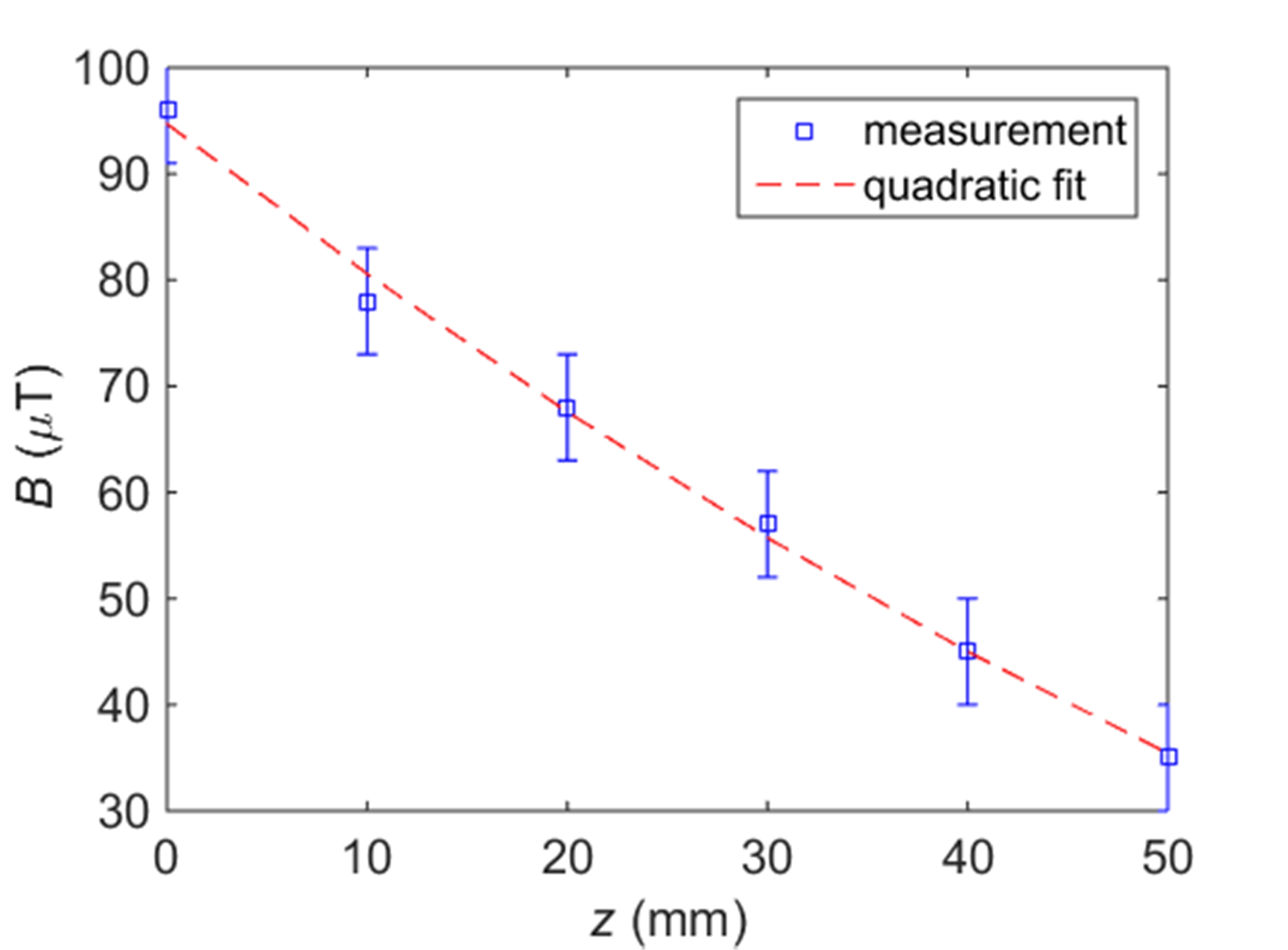}
\caption{The absolute magnitude of the magnetic field at the location of the mass. The measurement is fitted by a quadratic function. }
\label{fig10}
\end{figure}	

\section{Conclusions}
The active electromagnetic damping device presented in this paper can be used in both the dynamic and static measurement modes of the joule balance. Five degrees of freedom of the suspended coil, i.e. the horizontal displacement $x$, $y$ and the rotation angles $\theta_x$, $\theta_y$, $\theta_z$, can be actively controlled by the damping device. The experimental results show that the active damping device well meets the design targets. The damping factor of the system is increased by a factor of 10 and the alignment is greatly improved by compensating for the non-vertical movement of the coil. Two potential systematic effects, i.e. the unwanted vertical force and the leakage magnetic field of the active damping device, are analyzed. It is shown that these effects are small and compensable within few parts in $10^8$. At present, the misalignment error is about several parts in $10^7$ due to the restriction of the narrow air gap and the laser alignment in the joule balance project. The presented active damping device in this case is powerful to reduce misalignment uncertainties.

\ifCLASSOPTIONcaptionsoff
  \newpage
\fi



%

\begin{IEEEbiography}[{\includegraphics[width=1in,height=1.25in,clip,keepaspectratio]{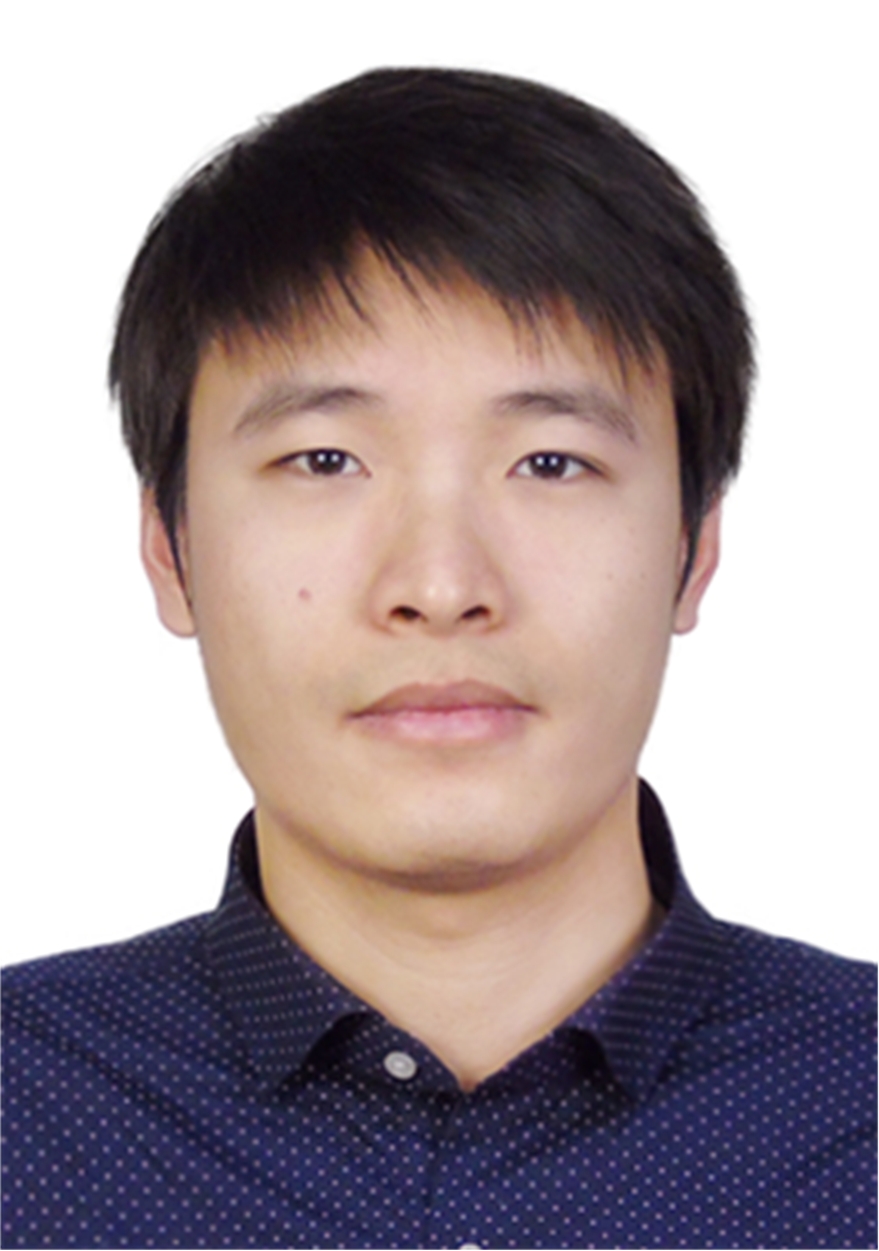}}]{Jinxin Xu}
 was born in Yancheng, Jiangshu province, China, in 1990. He received the B.S. degree from Harbin Institute of Technology, Harbin, China, in 2012. He is currently pursuing the Ph.D. degree at Tsinghua University, Beijing, China. His dissertation will be a part of the joule Balance at the National Institute of Metrology, China.
\end{IEEEbiography}

\begin{IEEEbiography}[{\includegraphics[width=1in,height=1.25in,clip,keepaspectratio]{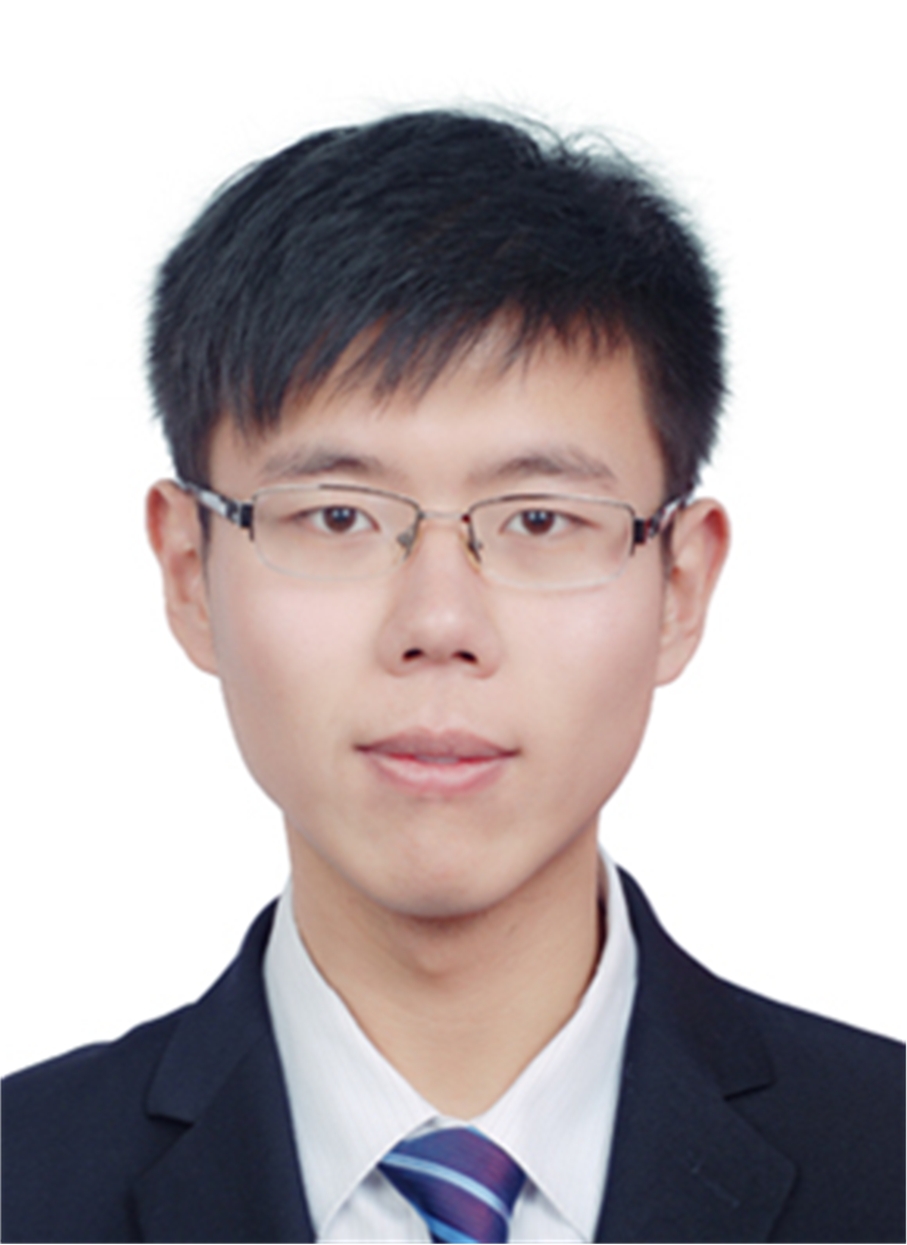}}]{Qiang You}
 was born in Shandong province, China. He is currently pursuing the Ph.D. degree with Tsinghua University, Beijing, China. His dissertation will be a part of the joule balance with the National Institute of Metrology, Beijing, China.
\end{IEEEbiography}

\begin{IEEEbiography}[{\includegraphics[width=1in,height=1.25in,clip,keepaspectratio]{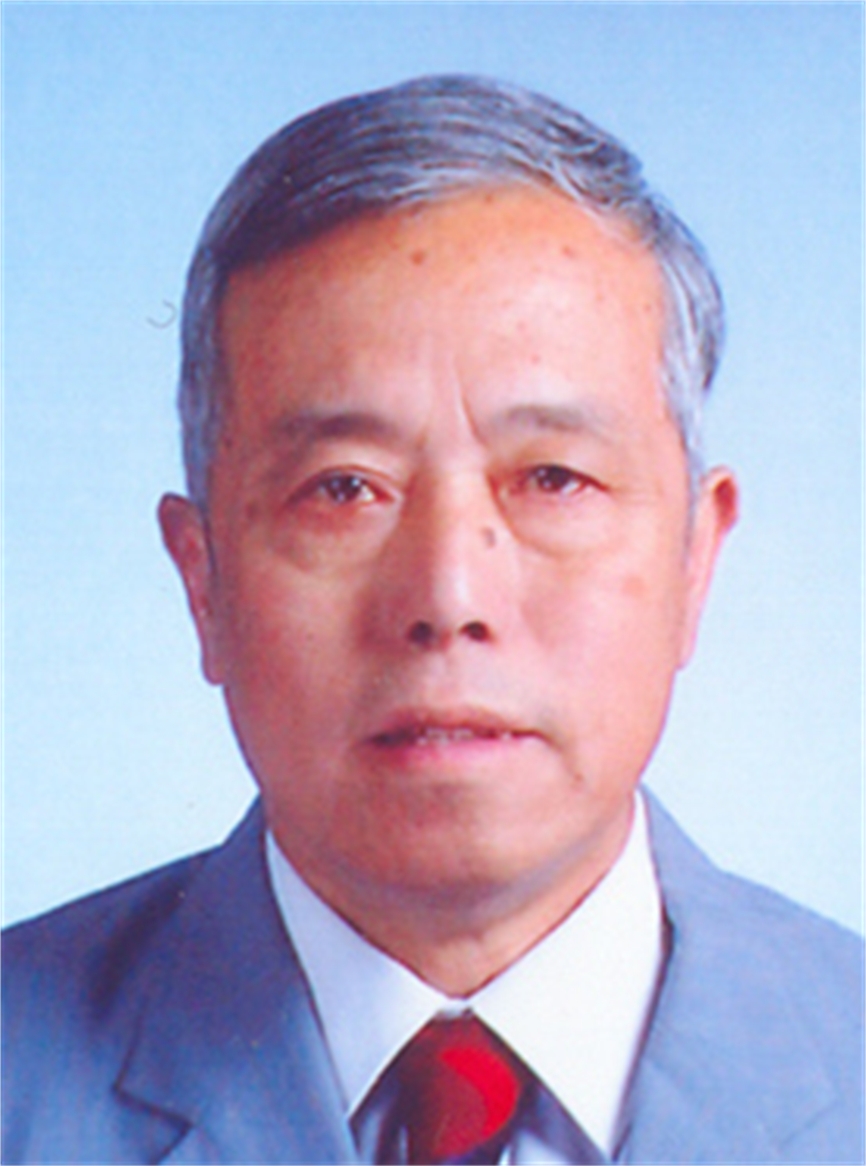}}]{Zhonghua Zhang}
was born in Suzhou, Jiangshu, China in July 1940. He received his B.S. degree and M.S. degree in Electrical Engineering from Tsinghua University, Beijing, China separately in 1965 and 1967. In 1995, he became a member of the Chinese Academy of Engineering.
In 1967, Zhonghua Zhang joined the Electromagnetic Division of National Institute of Metrology (NIM), Beijing, China. Since then he has been involved in the research work on the Cross Capacitor standard, superconducting high magnetic field standard and Quantum Hall Resistance standard. In 2006, he proposed the joule balance and led the research work since then.
\end{IEEEbiography}

\begin{IEEEbiography}[{\includegraphics[width=1in,height=1.25in,clip,keepaspectratio]{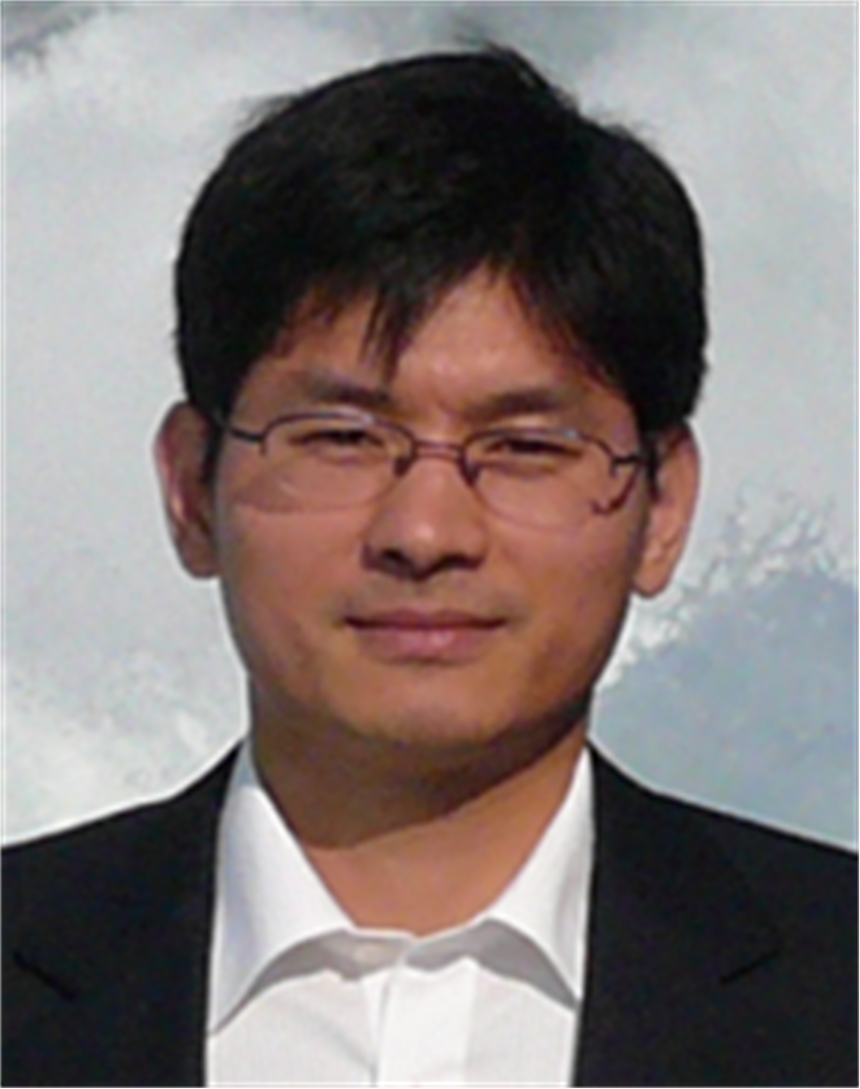}}]{Zhengkun Li}
was born in Henan Province, China in 1977. He received the B.S. degree in instrument science and technology from China Institute of Metrology, Hangzhou, China, in 1999. He received the M.S. degree in quantum division of National Institute of Metrology (NIM), China in 2002. He received his Ph.D. degree from Xi'an Jiaotong University, Xi'an, China in 2006.
In 2002, he became a permanent staff of NIM and worked on the establishment of Quantum Hall Resistance standard. Since 2006, he has been involved in the research work on joule balance project at NIM and focuses on the electromagnetic measurement including mutual inductance measurement for the joule balance.
\end{IEEEbiography}

\begin{IEEEbiography}[{\includegraphics[width=1in,height=1.25in,clip,keepaspectratio]{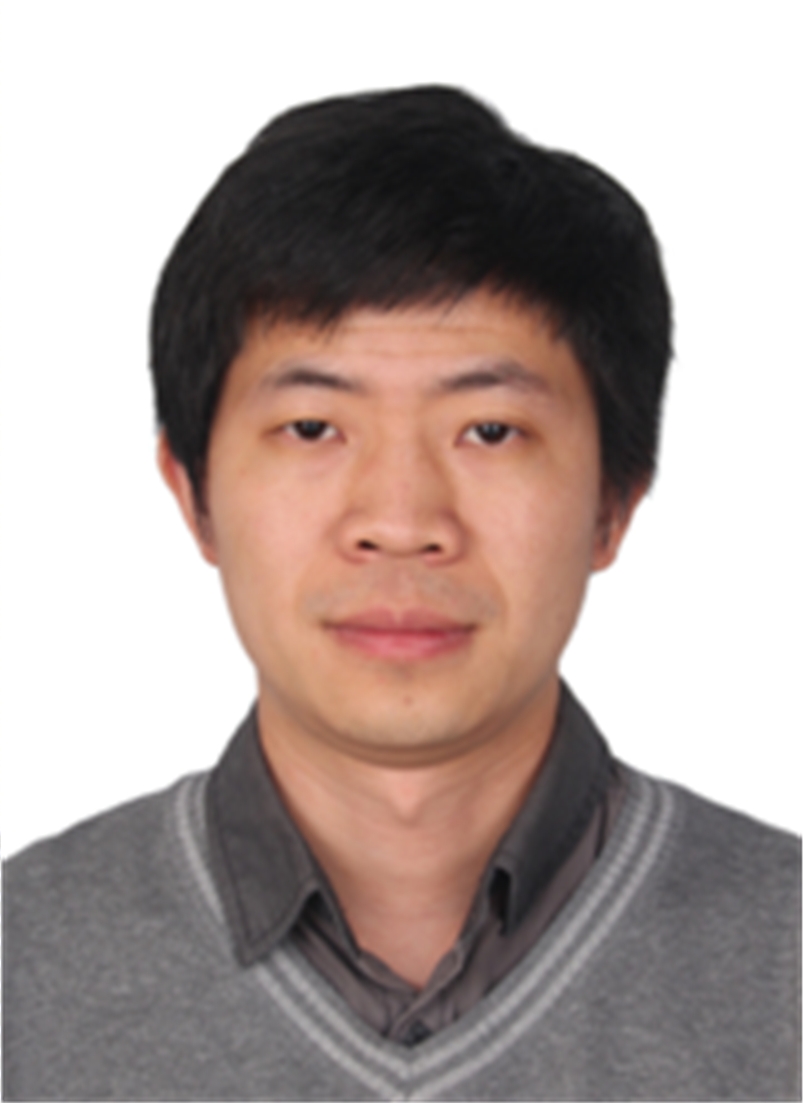}}]{Shisong Li}
  (M'15) got the Ph.D. degree in Tsinghua University, Beijing, China in July, 2014. He then joined the Department of Electrical engineering, Tsinghua University during July, 2014-September, 2016. He has been a guest researcher at the National Institute of Metrology in China since 2009, and he worked at the National Institute of Standards and Technology, United States for 14 months during 2013-2016. He is currently with the International Bureau of Weights and Measures (BIPM), France. His research interests include modern precision electromagnetic measurement and instrument technology.
\end{IEEEbiography}




\end{document}